\documentclass{PoS}

\newcommand\beq{\begin{equation}}
\newcommand\eeq{\end{equation}}
\newcommand\beqa{\begin{eqnarray}}
\newcommand\eeqa{\end{eqnarray}}

\newcommand\btau{\mbox{\boldmath$\tau$}}

\newcommand\bp{{\bf p}}
\newcommand\bk{{\bf k}}
\newcommand\br{{\bf r}}
\newcommand\bq{{\bf q}}
\newcommand\bx{{\bf x}}

\title{Fluctuations In The Inhomogeneous Chiral Transition}

\ShortTitle{Inhomogeneous Chiral Transition}

\author{\speaker{Toshitaka Tatsumi}\\ %\thanks{A footnote may follow.}\\
        Department of Physics, Kyoto University, Kyoto 606-8502, Japan\\
        E-mail: \email{tatsumi@ruby.scphys.kyoto-u.ac.jp}}

\author{Ryo Yoshiike\\
        Department of Physics, Kyoto University, Kyoto 606-8502, Japan\\
       E-mail: \email{yoshiike@ruby.scphys.kyoto-u.ac.jp}}

\author{T.-G. Lee\\
        Department of Physics, Kyoto University, Kyoto 606-8502, Japan\\
       E-mail: \email{tonggyu.lee@yukawa.kyoto-u.ac.jp}}

\abstract{Chiral pair fluctuation are considered near the phase boundary of the inhomogeneous 
chiral phase (iCP). The fluctuations are then bosonized and an effective action for the chiral pair fluctuation is basically constructed by considering the ring diagram of the polarization function. We can evaluate the self-energy and effective four point interaction among fluctuations in a consistent way. The peculiar dispersion of the fluctuation, reflecting the spatially inhomogeneous transition, gives rise to interesting and qualitative results. Thermal fluctuations prohibit the second-order transition, while the effect of the quantum fluctuations is rather modest. Quantum and thermal fluctuations changes the second-order transition to the first one by changing the sign of the effective four-point interaction between effective mesons. 
These features may be observed by relativistic heavy-ion collisions through the analysis of the thermodynamic observables. Distinct from the second-order phase transition, the first moment such as entropy production exhibits an anomalous behavior due to fluctuations, which is one of the signals of the phase transition to iCP.
Some similar aspects are also remarked between iCP and the FFLO state in superconductivity.}

\FullConference{The 26th International Nuclear Physics Conference\\
		11-16 September, 2016\\
		Adelaide, Australia}

\begin{document}

\section{Introduction}

There are many studies about the QCD phase diagram theoretically or experimentally on the temperature ($T$) and chemical potential ($\mu$) plane, where one may expect various phase transitions such as deconfinement transition or chiral transition. For the latter case, the system stays in the spontaneously symmetry broken (SSB) phase in the low $\mu$ and $T$ region, while chiral symmetry should be restored at high $\mu$ or $T$ region.
Recently a possibility of the spatially inhomogeneous chiral phase has been suggested and extensively studied in the various situations or within various theoretical frameworks \cite{tat,nic,bub}. Inhomogeneous chiral phase (iCP) is specified by the generalized quark condensate,
\beqa
M&=&\langle{\bar q}q\rangle+i\langle{\bar q}i\gamma_5\tau_3 q\rangle\nonumber\\
&=&\Delta(\br){\rm exp}\left(i\theta(\br)\right),
\eeqa
where amplitude $\Delta(\br)$ or phase $\theta(\br)$ is spatially modulating. Note that it includes the pseudo-scalar condensate as well as the usual scalar one. An example of the condensate may be the dual chiral density wave (DCDW), which describes the one dimensional modulation with $\Delta={\rm const.}, \theta=\bq_c\cdot\br$ \cite{tat}. The diagram of iCP has been studied within the mean-field approximation (MFA), using the effective models of QCD such as the NJL model. Fig.1 shows the phase diagram on the $T-\mu$ plane for DCDW: there appear three phases (SSB, chiral-restored and iCP phases ) separated by the two boundaries (denoted by $L$ and $R$) and these phase meet at the triple point called the Lifshitz point (LP). We can see the similar phase diagram for other cases like the real kink crystal (RKC) \cite{nic}, which is specified by the real and one dimensionally modulating condensate, $\theta=0, \Delta(\br)\in {\bf R}$. The effects of the magnetic field and the topological features of iCP 
 have been also discussed~\cite{fro,tatb,yos1,yos2}.

However, there are few studies about the fluctuation effects around the condensate. The fluctuations should consist of quark-antiquark pairs or quark particle-hole pairs and bear a bosonic nature. Two of the authors (T.T. and T.-G. L) have discussed the stability of the one-dimensional structure in iCP against the Nambu-Goldstone excitation \cite{lee}. These authors have found that the long range order is washed away by the thermal excitation of the Nambu-Goldstone modes, but the quasi-long-range order is realized instead. The correlation function in iCP has been shown to be algebraically decaying at large distance, which may imply that iCP could be considered to be realized in a realistic situation.

\begin{figure*}[h]
 \centering
   \includegraphics[width=2.5in]{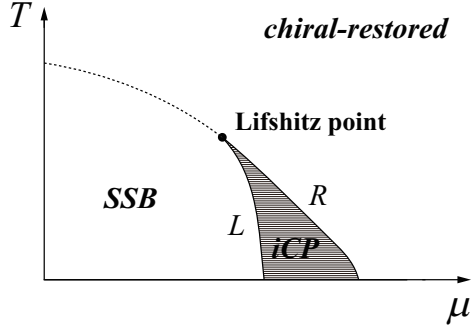}
 \caption{Schematic phase diagram
 for chiral transition in the $(\mu,T)$ plane.
The shaded domain 
 represents the inhomogeneous chiral phase (iCP).
 The $L$-boundary is of first or second order,
 depending on the type of iCP,
 whereas the $R$-boundary is always of second order.
}
~\\
 \label{fig1}
\end{figure*}

Here, we discuss the fluctuation effects near the phase boundary. iCP is surrounded by the $L$ and $R$ boundaries. The $L$ boundary depends on the definite form of the condensate, but the $R$ boundary should have a common feature to various condensates; there should be no distinction because the restored phase is chirally symmetric and we cannot see what type of the condensate is realized after phase transition. Actually it has been shown the $R$ boundary is the same for both DCDW and RKC \cite{car}. The $R$ boundary indicates the second order phase transition within the mean-field approximation. Thus we discuss the effect of the fluctuations near the $R$ boundary in this talk \cite{kar,yos}.

\section{Brazovskii and Dyugaev effect}

We start with the partition function $Z$ within the two flavor NJL model in the chiral limit to discuss the fluctuation effects in a systematic way \cite{kar},
\beqa
Z&=&\int{\cal D}q{\cal D}{\bar q}e^{-S}, \nonumber\\
S&=&-\int_0^\beta d\tau\int d^3x\left[{\bar q}\left(-\gamma_0\frac{\partial}{\partial\tau}\right)q+G\left(({\bar q}q)^2+({\bar q}i\gamma_5\btau q)^2\right)\right].
\eeqa
Here we introduce the auxiliary fields written by the quark bilinear fields, $\phi_a=-2G({\bar q}q,{\bar q}i\gamma_5\btau q)$. After integrating out the quark variable we have an effective action $S_{\rm eff}$ in terms of $\phi_a$,
\beq
S_{\rm eff}\sim \int_0^\beta d\tau\int d^3x \left[\frac{1}{2!}\Gamma^{(2)}\phi^2_a+\frac{1}{4!}\Gamma^{(4)}(\phi_a^2)^2+...\right],
\eeq
and $S_{\rm eff}$ respects $SU(2)_L\times SU(2)_R\sim O(4)$ symmetry. Each coefficient $\Gamma^{(n)}$ denotes the bare vertex function for $\phi_a$ and can be evaluated within MFA. Separating $\phi_a$ into the condensate and the chiral pair fluctuations around it, $\phi_a=\Phi(\Delta,q)+\xi_a$, we have the thermodynamic potential,
%\end{document}
\beqa
\Omega-\Omega_f=&&T\sum_{q,\omega_n}{\rm log}\left(1-2G{\bar\Pi}_{ps}^0(q,\omega_n)\right)+\nonumber\\
&+&\int d^3x\left[\frac{1}{2!}{\bar \Gamma}^{(2)}\left|\Phi(\bx)\right|^2+\frac{1}{4!}{\bar \Gamma}^{(4)}\left|\Phi(\bx)\right|^4+\frac{1}{6!}{\bar \Gamma}^{(6)}\left|\Phi(\bx)\right|^6+...\right],
\eeqa
%\end{document}
with the Matsubara frequency, $\omega_n=2n\pi T$.
The coefficients ${\bar\Gamma}^{(n)}$ are the renormalized vertex functions and include the fluctuation effects. The first term comes from the chiral pair fluctuation in the chiral-restored phase and is given by the ring diagrams composed of the q-anti q and particle-hole polarization function ${\bar\Pi}^0_{ps}$. The second term tell us the shape change of the effective potential as a function of the condensate (Fig.~2). Note that the first term is very similar to the superconducting phase transition in the context of the Noziere and Schmitt-Rink theory \cite{nsr}. The second term implies the non-linear effects beyond the Gaussian approximation.

\begin{figure*}[h]
 \centering
   \includegraphics[width=2.5in]{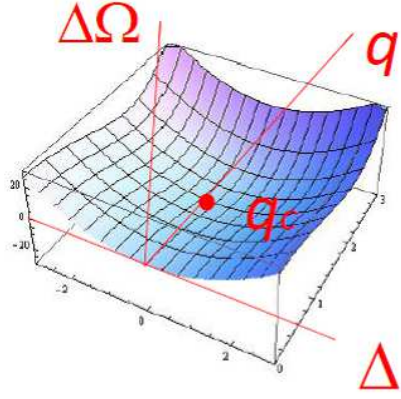}
 \caption{Effective potential near the phase boundary in the chiral-restored phase. The minimum point is located at $(\Delta,q)=(0,q_c)$.
}
~\\
 \label{fig1}
\end{figure*}

We look into the polarization function ${\bar\Pi}^0_{ps}$ in detail. The bare function is given as 
\beq
\Pi^0_{ps}(i\nu_n,\bq)=-\sum_{m,\bp}{\rm tr}\left[i\gamma_5\tau_3S_\beta(p+q)i\gamma_5\tau_3S_\beta (p)\right],
\eeq 
with $p_0=i(2m+1)\pi T+\mu, q_0=i2n\pi T$, where $S_\beta$ is the thermal propagator for massless quarks. Then we find that the static polarization function exhibits an interesting feature; when $\partial^2\Pi^0_{ps}(0,\bq)/\partial q^2|_{q=0}<0$, it has an extremum at non-zero $q_c$. We shall see that such behavior gives rise to remarkable effects. The bare Green function  $G_{ps}$ for the chiral pair field $\xi_a$ then can be constructed by summing the bubble diagrams made of $\Pi_{ps}^0$,
\beq
G_{ps}(i\nu_n,\bq)=\left(1-2G\Pi_{ps}^0(i\nu_n,\bq)\right)^{-1},
\eeq
which behaves $G_{ps}(i\nu_n,\bq)^{-1}\sim \tau+\gamma\left(|\bq|^2-q_c^2\right)^2+\alpha|\nu_n|$ near the phase boundary. One may then read a peculiar dispersion of the chiral pair fluctuations through the pole of $G_{ps}$
\footnote{In the usual case for the homogeneous phase transitions, the dispersion of the fluctuations should exhibit the form, $\tau+\gamma q^2$, near the critical point.}
. 
Since the sign change of the effective potential at the origin is the signal of the second order phase transition, we can examine the static Green's function for the critical point: the singular point of $G_{ps}(0,\bq)$ resides on $\tau=0,|\bq|=q_c$ and corresponds to the Thouless criterion \cite{tho} within MFA. Thus fluctuations become soft with momentum on the two-dimensional sphere $|\bq|=q_c$ near the phase boundary.

Once the thermodynamic potential is evaluated by including the fluctuation effects, the renormalized $n-$th order vertex function can be obtained through the relation,
\beq
{\bar\Gamma}^{(n)}=(2\pi)^{3n}\left.\frac{\delta^n\Omega}{\delta\Phi(-\bq_1)\delta\Phi(-\bq_2)...\delta\Phi(-\bq_n)}\right|_{\Phi=0},
\eeq
which is reduced to the bare vertex function $\Gamma^{(n)}$ within MFA. Diagramatic expansion of $\Omega$ can be done for small $\Gamma^{(4)}$ by using Eq.~(2.2). In the following we replace the non-local $\Gamma^{(4)}$ by the local one, $\lambda$, for simplicity. First, we consider the second-order vertex function, which can be obtained by way of the Dyson equation: the $\tau$ parameter in $G_{ps}$ is just replaced by the renormalized value $\tau_R$,
\beq
{\bar\Gamma}^{(2)}(\bq_1,\bq_2)\propto \delta(\bq_1+\bq_2)\left(G_{ps}^R(i\nu_n,\bq_1)\right).
\eeq
Then we have a relation between $\tau$ and $\tau_R$,
\beqa
\tau&=&\tau_R-\lambda T\sum_{\nu_n}\int\frac{d^3p}{(2\pi)^3}G_{ps}^R(i\nu_n,\bp)\nonumber\\
&\simeq&\tau_R-\frac{\lambda T q_c}{4\pi\gamma^{1/2}\tau_R^{1/2}},~~~T\neq 0\\
&\simeq&\tau_R-\frac{\lambda\Lambda}{48\alpha\pi^{5/2}\gamma^{3/2}q_c^3},~~~T=0,
\eeqa
with an ultraviolet cut-off $\Lambda$. Here we can see a remarkable effect of the thermal fluctuations:  $\tau_R$ is never vanished from Eq.~(2.8), which implies that the second order phase transition is prohibited by the thermal fluctuations. On the other hand, the quantum fluctuations have a gentle effect by shifting the critical point from MFA (Fig.3).

\begin{figure*}[h]
 \centering
   \includegraphics[width=2.5in]{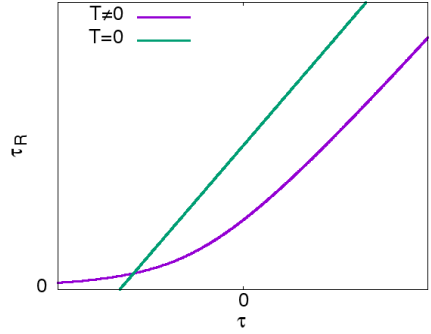}
 
\caption{$\tau_R$ vs $\tau$ for $T=0$ and $T\neq 0$ cases. $\tau_R$ never reaches 0 to prohibit the second order phase transition by the thermal fluctuation.
}
~\\
 \label{fig2}
\end{figure*}

We can see how such difference is produced for quantum and thermal fluctuations by looking into the one loop integral for $\tau_R$. All the Matsubara frequencies contribute to the integral at $T=0$, which washed out the singularities included in $G_{ps}^R$ in the integral. On the other hand, $n=0$ gives a leading contribution at finite temperature,
\beq
T\sum_{\omega_n}\int\frac{d^3p}{(2\pi)^3}\rightarrow T\int\frac{d^3p}{(2\pi)^3}.
\eeq
Thus we can see there occurs {\it dimensional reduction}, so that the momentum integral in the lower dimension becomes singular.

Next we examine the forth-order vertex function. It is obtained by the infinite sum of the bubbles  $L(\bk)$ made of the two Green's function. $L(\bk)$ gives the long-range interaction between vertexces in the coordinate space,
\beq
{\bar\Gamma}^{(4)}=(2\pi)^3\lambda\frac{1-\frac{1}{3}\lambda L(0)}{1+\lambda L(0)}.
\eeq
$L(\bk)$ exhibits a singularity at $\bk=0$, since the singularities included in two Green's function come closer together as $k\rightarrow 0$,
\beqa
L(0)&=&\lim_{k\rightarrow 0}T\sum_{\omega_n}\int\frac{d^3p}{(2\pi)^3}G_{ps}^R(i\omega_n,\bp)G_{ps}^R(-i\omega_n,\bk-\bp)\nonumber\\
&\simeq&\frac{T q_c}{8\pi\gamma^{1/2}}\frac{1}{\tau_R^{3/2}},~~~T\neq 0\\
&\simeq&\frac{q_c}{4\alpha\pi^2\gamma^{1/2}}\frac{1}{\tau_R^{1/2}},~~~T=0.
\eeqa
Thus ${\bar\Gamma}^{(4)}$ changes its sign as $\tau_R\rightarrow 0$, which sgnals the first order phase transition. This is called the Brazovskii-Dyugaev effect \cite{bra,dyu}.

Since ${\bar\Gamma}^{(6)}$ is always positive, we can trace the change of the thermodynamic potential on the $\mu-T$ plane.

\section{Anomalies of the thermodynamic quantities}

Finally, we give a brief argument about the phenomenological implications of the fluctuation induced first order phase transition. We shall see anomalies of the thermodynamic quantities 
near the fluctuation induced first order phase transition.

For the second order phase transition many authors have discussed the various susceptibilities or second derivatives of the thermodynamic potential near the phase transitions; specific heat or number susceptibility may be a typical example \cite{fuj}. For example heat capacity diverges like $(T-T_c)^{-1/2}$ due to the fluctuations near the superconducting phase transition. For the fluctuation induced phase transition we may expect anomalies in the first derivatives of the thermodynamic potential instead: number density or entropy density is suitable \cite{oha}. Considering the contribution of the chiral pair fluctuations,
\beq
\Delta\Omega=2TV\sum_{\bq,\omega_n}{\rm ln}\left(1-2G{\bar\Pi^0_{ps}}(i\omega_n,\bq)\right),
\eeq
Accordingly, entropy density, for example, reads \cite{yos}
\beq
\Delta s=- \frac{1}{\tau_R^{1/2}} \frac{N_fN_cq_c}{8\pi^3 \gamma^{1/2}T} \int_{-\infty}^\infty dp \frac{p(p+\mu)}{\left[ e^{\beta(p+\mu)/2} + e^{-\beta(p+\mu)/2} \right]^2} \left( 4 + \frac{q_c}{p} \ln\left| \frac{2p-q_c}{2p+q_c}\right| \right).
\eeq
Such anomalous behavior may be observed in the relativistic heavy-ion collisions through the multiplicity of the particle production.

\section{Summary and concluding remarks}

We have discussed the effects of the chiral pair fluctuations on the inhomogeneous chiral transition. 
We have taken into account the non-linear effects of the chiral-pair fluctuations in a systematic way, beyond the Gaussian approximation.
We have elucidated the salient roles of the quantum and thermal fluctuations separately:
the latter is more drastic than the former due to the dimensional reduction, but both lead to the fluctuation induced first-order phase transition.
The curvature parameter $\tau$ is renormalized by the fluctuation effects to be positive definite at $T\neq 0$, while it is mildly shifted from the one within MFA at $T=0$. Thus we have observed that the second order phase transition is prohibited by the thermal fluctuations. More importantly, the dangerous diagrams composed of the bubbles of two fluctuation Green's function become essential  and change the sign of the fourth-order vertex function for both cases at $T=0$ and finite temperature. The sign of the sixth-order vertex function can be shown to be positive definite and we can clearly see the first-order phase transition. These features are brought about by the unique behavior of the dispersion of the chiral pair fluctuations and common in any inhomogeneous phase transition.

Note that these particular roles of the fluctuations have some similarities to those of the Nambu-Goldstone excitations in iCP \cite{yos}: the thermal excitations of the Nambu-Goldstone modes lead to the instability of the one dimensional configuration (the Landau-Peierls theorem \cite{lan,pei}), while it is stable against the quantum excitations.  

It should be worth mentioning that the behavior of the vertex functions has been also studied by solving the flow equations within the renormalization group approach, and the findings with the perturbative approach have been confirmed for the diblock copolymer \cite{hoh}. The renormalization group is somewhat different from the usual one due to the existence of the special point $q_c$ in the momentum space, but can be formulated in the similar way to the work by Shankar for the fermion many-body system \cite{sha}, where the Fermi momentum corresponds to $q_c$.
Since our formalism is very much similar to theirs, one may expect that our findings are also confirmed by the renormalization group approach. This is left for a future work. 

The first derivative of the thermodynamic potential exhibits a singular behavior through the momentum integral, since the dispersion of chiral-pair fluctuations has a minimum on the sphere, $|\bq|=q_c$. We have evaluated the number density and entropy density to figure out singular behavior.
Thus the fluctuation-induced phase transition can be characterized by the discontinuity and singular behavior of the first derivatives. 

Finally it should be worth mentioning that our formalism to treat the non-linear effects of the fluctuations may be also applied to the other cases like the FFLO state in superconductivity;
the Cooper pair fluctuations are composed of the particle-particle ladder diagram instead, but the dispersion relation has a similar feature discussed here. Accordingly the entropy anomaly may be a possible evidence for the phase transition.

\section*{Acknowledgement}

We thank Y. Ohashi for stimulating discussions.
This work is supported by the Grants-in-Aid for Japan Society for the Promotion of Science (JSPS) fellows No.~27-1814 and in part by the Grant-in-Aid for Scientific Research on Innovative Areas through No. 24105008 provided by MEXT.

\end{document}